\documentstyle[prl,aps,twoside,multicol]{revtex}

\begin{document}
\title{Power of Anisotropic Exchange Interactions: Universality and
Efficient Codes for Quantum Computing}
\author{L.-A. Wu and D. A. Lidar}
\address{Chemical Physics Theory Group, University of Toronto, 80
St. George St., Toronto, Ontario M5S 3H6, Canada}
\maketitle

\begin{abstract}
We study the quantum computational power of a generic class of anisotropic
solid state Hamiltonians. A universal set of encoded logic operations are
found which do away with difficult-to-implement single-qubit gates in a
number of quantum computer proposals, e.g., quantum dots and donor atom
spins with anisotropic exchange coupling, quantum Hall systems, and
electrons floating on helium.We show how to make the corresponding
Hamiltonians universal by encoding one qubit into two physical qubits, and
by controlling nearest neighbor interactions.
\end{abstract}

\begin{multicols}{2}

While decoherence is the most significant fundamental obstacle in the path
towards the construction of a quantum computer (QC), in the realm of
scalable QC proposals \cite
{Loss:98,Kane:98,Vrijen:00,Privman:98,Imamoglu:99,Zheng:00,Platzman:99} a
pressing concern is the technological difficulty of implementing
single-qubit operations together with two-qubit operations. In general these
two types of operations may impose very different constraints, or
single-qubit operations may be hard. E.g., in the proposals utilizing
quantum dots \cite{Loss:98}, donor-atom nuclear \cite{Kane:98} or electron 
\cite{Vrijen:00} spins, and quantum Hall systems \cite{Privman:98},
single-qubit operations require control over a local magnetic field, are
significantly slower than two-qubit operations (mediated by an exchange
interaction), and require substantially greater materials and device
complexity. In the quantum dots in cavities proposal \cite{Imamoglu:99} each
dot needs to be illuminated with a separate laser, and reduction in the
number of lasers by elimination of single-qubit operations is a potentially
significant technical simplification. In the electrons-on-helium proposal 
\cite{Platzman:99} single-qubit operations require slow microwave pulses,
limiting the number of logic operations executable before decoherence sets
in. It is thus clear that quite generally a significant gain may be had by
enabling quantum logic operations to be performed through two-qubit
operations only. The need for single-qubit operations arises from the
``standard paradigm'' of (non fault-tolerant) universal quantum computation,
which prescribes the use of single-qubit Hamiltonians that can generate all
one-qubit quantum gates [$SU(2)$] together with a two-body interaction that
can generate an entangling two-qubit gate such as {\sc CNOT} \cite
{Barenco:95a}. The universality of this set essentially amounts to its
ability to generate $SU(2^{N})$ with $N$ qubits \cite{Nielsen:book}. While
it was recognized early on that\ a universal QC can be constructed using at
most two-body interactions \cite{Lloyd:95}, the abstract theory hardly makes
reference to the ``natural talents'' of a given quantum system as dictated
by its intrinsic Hamiltonian. Indeed, most discussions of universality,
e.g., \cite{Gottesman:97a}, rather than using the physical notion of
Hamiltonians, are cast in the computer-science language of unitary gates
(exponentiated Hamiltonians). Based on these observations a new paradigm was
recently proposed in \cite{Bacon:Sydney}, termed ``encoded-universality''
(EU): to study the quantum computational power of a system {\em as embodied
in its naturally available Hamiltonian}, by using encoding [encoded gates --
consisting of sequences of physical gates -- act on encoded (logical) qubits
generating $SU(2^{M})$, where $M$ is the dimension of the code space].
Earlier work \cite
{Bacon:99a,Zanardi:99dLidar:00bViola:01,Kempe:00,DiVincenzo:00a} had
implicitly studied EU constructions. In this work we introduce a general
formalism, discovered by a mapping of qubits to parafermions that will be
described elsewhere \cite{WuLidar:tbp}, that allows us to quickly assess the
quantum computational power of a given Hamiltonian, and construct encoded
qubits and operations. Our main result is the classification of the EU power
of generic classes of solid-state Hamiltonians, addressing in particular the
case of {\em anisotropic} qubit-qubit interactions pertinent to the quantum
Hall \cite{Privman:98}, quantum dots \cite{Imamoglu:99} and atoms \cite
{Zheng:00} in cavities, and the electrons-on-helium \cite{Platzman:99}
proposals. The proposals relying on purely isotropic (Heisenberg)\ exchange
may also benefit from our analysis, in the case that some symmetry breaking
mechanism (e.g., surface and interface effects, and/or spin-orbit coupling 
\cite{Kavokin:00}) introduces anisotropy. For all these cases we give
explicit EU constructions which avoid the use of the undesirable
single-qubit gates. In particular, we show how to make the anisotropic
exchange Hamiltonian universal by {\em encoding one qubit into two physical
qubits}, in contrast to previous results for the Heisenberg case where three
physical qubits were required \cite{Bacon:Sydney,Kempe:00,DiVincenzo:00a}.
Only nearest-neighbor couplings are needed in this construction. Thus we
suggest new ways to simplify the operation of a variety of QC proposals,
circumventing operations that appear to be dictated by the ``standard
paradigm''.

{\it General analysis.} --- To set the stage for our discussion of the
universality properties of Hamiltonians, let us consider the general
structure of operators in the Hilbert space of $N$ qubits in terms of the
lowering and raising operators $\sigma _{i}^{\pm }=(\sigma _{i}^{x}\mp
i\sigma _{i}^{y})/2$, where $i=1,...,N$ and $\sigma _{i}^{\alpha }$ acts
non-trivially only on the $i^{{\rm th}}$ qubit. Define an occupation number $
n_{i}=(1-\sigma _{i}^{z})/2=0$ or $1$, which is the number of $1$'s
(up-spins) in the $i^{{\rm th}}$ position of the vectors of the
computational basis, i.e., all length-$N$ bitstrings. The most general
operator consistent with $\sigma _{i}^{-}\sigma _{i}^{-}=\sigma
_{i}^{+}\sigma _{i}^{+}=0$ is a linear combination of

\begin{equation}
Q_{\{\alpha \}\{\beta \}}=(\sigma _{N}^{+})^{\alpha _{N}}\cdots (\sigma
_{1}^{+})^{\alpha _{1}}(\sigma _{N}^{-})^{\beta _{N}}\cdots (\sigma
_{N}^{-})^{\beta _{1}}
\end{equation}
where $\alpha _{i},\beta _{j}$ can be $0$ or $1$. There are $2^{N}\times
2^{N}$ such operators which form a complete set of generators of the group $
U(2^{N})$ needed for universal quantum computing\cite{comment1}. They can be
rearranged into certain subsets of operators with clear physical meaning,
which we now detail. First, there is a subalgebra with conserved total
occupation number, ``SA$n$''. This is formed by all operators commuting with
the total number operator $\widehat{n}=\sum_{i}n_{i}$. Let $k$ ($l$) be the
number of $\sigma _{i}^{+}$ ($\sigma _{i}^{-}$) factors in $Q_{\{\alpha
\}\{\beta \}}$. SA$n$ consists of the operators for which $k=l$, so the
dimension of SA$n$ is $\sum_{n=0}^{N}{
{N \choose n}}^{2}=\frac{(2N)!}{N!N!}$. Second, there is a subalgebra with conserved
parity, ``SA$p$'', i.e., the operators commuting with the parity operator,
defined as $\widehat{p}=(-1)^{\widehat{n}}$, with eigenvalues $1$ ($-1$) for
even (odd)\ total occupation number. SA$p$ consists of those operators
having $k-l$ even, so its dimension is $2^{2N}/2$. Clearly, SA$n$$\subset $
SA$p$. Third, there are types of $su(2)$ subalgebras generated by the set $
\{Q_{\{\alpha \}\{\beta \}},Q_{\{\alpha \}\{\beta \}}^{\dagger
},[Q_{\{\alpha \}\{\beta \}},Q_{\{\alpha \}\{\beta \}}^{\dagger }]\}$ in the
subspace satisfying the condition $\{Q_{\{\alpha \}\{\beta \}},Q_{\{\alpha
\}\{\beta \}}^{\dagger }\}=1$, for specific choices of $\{\alpha \}\{\beta \}$. This results directly in encoding schemes. The following two types of
bilinear operators for $i\neq j$: $\sigma _{i}^{+}\sigma _{j}^{-}$ (which
conserve the occupation number), and $\sigma _{i}^{-}\sigma _{j}^{-},\sigma
_{i}^{+}\sigma _{j}^{+}$ (which conserve parity), are important examples
that illustrate this case. Let $\mu =(ij)$, then 
\begin{equation}
T_{\mu }^{x}=\sigma _{j}^{+}\sigma _{i}^{-}+\sigma _{i}^{+}\sigma _{j}^{-}
\text{ and }T_{\mu }^{z}=\sigma _{i}^{z}-\sigma _{j}^{z}  \label{eq:slt}
\end{equation}
generate an $su(2)$ subalgebra, denoted $su_{\mu }^{t}(2)\in $SA$n$.

\begin{equation}
R_{\mu }^{x}=\sigma _{i}^{-}\sigma _{j}^{-}+\sigma _{i}^{+}\sigma _{j}^{+} 
\text{ and }R_{\mu }^{z}=\sigma _{i}^{z}+\sigma _{j}^{z}  \label{eq:slr}
\end{equation}
generate another $su(2)$ subalgebra, denoted $su_{\mu }^{r}(2)\in $SA$p$. It
is easy to show that $[su_{\mu }^{t}(2),su_{\mu }^{r}(2)]=0$. It can be
shown that $\{\sigma _{i}^{+}\sigma _{j}^{-}\}$ (allowing $i=j$) generates SA
$n$, and $\{\sigma _{i}^{+}\sigma _{j}^{-},\sigma _{i}^{-}\sigma
_{j}^{-},\sigma _{i}^{+}\sigma _{j}^{+}\}$ generate SA$p$ \cite{WuLidar:tbp}.

{\it Hamiltonians and universal sets without single-qubit operations}.---
Now consider the properties of Hamiltonians relevant to scalable proposals
for quantum computing. A generic time-dependent Hamiltonian \cite
{Loss:98,Kane:98,Vrijen:00,Privman:98,Imamoglu:99,Zheng:00,Platzman:99,Nielsen:book}
has the form 
\begin{eqnarray}
H(t) &\equiv &H_{0}+V+F  \nonumber \\
&=&\sum_{i}\frac{1}{2}\varepsilon _{i}(t)\sigma
_{i}^{z}+\sum_{i<j}\sum_{\alpha ,\beta =x,y,z}J_{ij}^{\alpha \beta
}(t)\sigma _{i}^{\alpha }\sigma _{j}^{\beta }  \nonumber \\
&&+\sum_{i}\left( f_{i}^{x}(t)\sigma _{i}^{x}+f_{i}^{y}(t)\sigma
_{i}^{y}\right) .  \label{eq:H(t)}
\end{eqnarray}
The first term is the sum of single-qubit energies, (with $\varepsilon
_{i}/\hbar $ being the frequency of the $|0\rangle _{i}\rightarrow |1\rangle
_{i}$ transition) and is often controllable using local potentials. The
second term is the two-qubit interaction, which we assume can be turned
on/off at controllable times $t$. The third term is the (potentially
problematic) external field, often pulsed, used to manipulate single qubits.
By turning the controllable parameters on/off one has access to a set of
Hamiltonians $\{H_{i}\}$, which can be used to generate unitary logic gates
through the following three processes: (i) {\em Arbitrary phases} are
obtained by switching an $H_{i}$ on for a fixed time. (ii) {\em Adding} or
(iii) {\em commuting} {\em Hamiltonians} can be approximated by using a
finite number of terms in the Lie sum and product formulas, e.g. \cite
{Nielsen:book,Lloyd:95}, $e^{i(\alpha A+\beta B)}=\lim_{n\rightarrow \infty
}\left( e^{i\alpha A/n}e^{i\beta B/n}\right) ^{n}$, implying that the
Hamiltonians $A$, $B$ are switched on/off alternately. These operations are
experimentally implementable and suffice to cover the Lie group generated by
the set $\{H_{i}\}$. In practice it may be easier to use Euler angle
rotations rather than infinitesimal steps \cite{DiVincenzo:00a}, as done
routinely in NMR\ {\cite{Nielsen:book}}. Let us now specialize to the case $
J_{ij}^{\alpha \beta }=J_{ij}^{\alpha }\delta _{\alpha \beta }$ (denoting $V$
by $V^{\prime }$) which amounts to limiting the Hamiltonian to exchange-type
interactions, that appear to be most relevant for solid-state QC. Using $
\sigma _{i}^{\pm },n_{i}$ we find

\begin{equation}
H_{0}=\sum_{i}\varepsilon _{i}(\frac{1}{2}-n_{i}),\;\;\;F=\sum_{i}\left(
f_{i}^{\ast }\sigma _{i}^{-}+f_{i}\sigma _{i}^{+}\right) ,  \label{eq:F}
\end{equation}

\begin{eqnarray}
V^{\prime } &=&\sum_{i<j}\left( \Delta _{ij}(\sigma _{i}^{-}\sigma
_{j}^{-}+\sigma _{i}^{+}\sigma _{j}^{+})+J_{ij}(\sigma _{i}^{+}\sigma
_{j}^{-}+\sigma _{j}^{+}\sigma _{i}^{-})\right.   \label{eq:V'} \\
&&+\left. J_{ij}^{z}\sigma _{i}^{z}\sigma _{i}^{z}\right) 
\end{eqnarray}
where 
\[
f_{i}=(f_{i}^{x}-if_{i}^{y}),\quad \Delta _{ij}=J_{ij}^{x}-J_{ij}^{y},\quad
J_{ij}=J_{ij}^{x}+J_{ij}^{y}.
\]
The above analysis of the subalgebras of $U(2^{N})$ now helps us in drawing
certain general conclusions. (i) By appending $\sigma _{i}^{-},\sigma
_{i}^{+}$ to the set generating SA$p$ it becomes possible to transform
between states differing by an odd occupation number. Thus the set $\{\sigma
_{i}^{+}\sigma _{j}^{-},\sigma _{i}^{-}\sigma _{j}^{-},\sigma _{i}^{+}\sigma
_{j}^{+},\sigma _{i}^{-},\sigma _{i}^{+}\}$ suffices to generate $SU(2^{N})$. This establishes the well-known universality of $H$. (ii) When $F=0$, $
[H_{0}+V^{\prime },\widehat{p}]=0$, so $H_{0}+V^{\prime }$ is in SA$p$. This
implies that this Hamiltonian by itself is {\em not fully universal}: it
operates on a $2^{N-1}$-dimensional invariant subspace. (iii) Recalling that
single qubit operations are often difficult, which two-qubit interactions
are sufficient for universality? Ref. \cite{Lloyd:95} established that
two-body Hamiltonians are ``generically'' universal. The genericness
condition was stated in terms of abstract group-theoretic properties. Here
we are able to state the condition more explicitly for the class of Eq.~(\ref
{eq:H(t)}). Define the parity of an operator according to whether the total
number of raising and lowering operators is even or odd (e.g., $n_{1}$ is
even, but $\sigma _{2}^{-}n_{1}$ is odd.). The necessary condition for a
Hamiltonian to be universal is that it contains an odd term, so that the
system can leave SA$p$. If $F=0$ there does not exist an odd term in $H(t)$.
Hence the next step is to reconsider the most general interaction with $
J_{ij}^{\alpha \beta }$ arbitrary$.$ $H$ of Eq.~(\ref{eq:H(t)}) is universal
for $F=0$ if and only if there exists one of the odd terms $\sigma
_{i}^{z}\sigma _{j}^{x}=(1-2n_{i})(\sigma _{j}^{+}+\sigma _{j}^{-})$ or $
\sigma _{i}^{z}\sigma _{j}^{y}$. Such terms may arise due to perturbative
spin-orbit coupling corrections to the isotropic part $J_{ij}(t)
\overrightarrow{\sigma }_{i}\cdot $ $\overrightarrow{\sigma }_{j}$ [where $
\overrightarrow{\sigma }_{i}=(\sigma _{i}^{x},\sigma _{i}^{y},\sigma
_{i}^{z})$] of Eq.~(\ref{eq:H(t)}). E.g., a recent estimate of the coupling
strength of the antisymmetric (Dzyaloshinskii-Moriya) spin exchange term $
\overrightarrow{d_{ij}}\cdot (\overrightarrow{\sigma }_{i}\times $ $
\overrightarrow{\sigma }_{j})$ shows $|\overrightarrow{d_{ij}}|/J_{ij}$ to
be as large as 0.01\ for coupled quantum dot in GaAs \cite{Kavokin:00}.
Unlike the isotropic exchange parameter $J_{ij}(t)$, $\overrightarrow{d_{ij}}
$ is typically {\em not }controllable. Nevertheless, its very presence
allows for universal QC without the external field $F$. To see this, suppose
for simplicity that $\overrightarrow{d_{ij}}$ is along the $x$-axis [so that 
$\overrightarrow{d_{ij}}\cdot $ $(\overrightarrow{\sigma }_{i}\times $ $
\overrightarrow{\sigma }_{j})=d_{ij}($ $\sigma _{i}^{y}\sigma _{j}^{z}-$ $
\sigma _{i}^{z}\sigma _{j}^{y})$], and that the terms $\overrightarrow{
\sigma }_{i}\cdot $ $\overrightarrow{\sigma }_{j}$, $\sigma _{i}^{z}$ are
controllable while $\sigma _{i}^{y}\sigma _{j}^{z}-$ $\sigma _{i}^{z}\sigma
_{j}^{y}$ is small and not controllable. Then we can show that these
operators generate the group $SU(4)$ on the qubit pair $i,j$ and therefore
are universal. The Hamiltonian is $H_{ij}=d_{ij}(\sigma _{i}^{y}\sigma
_{j}^{z}-\sigma _{i}^{z}\sigma _{j}^{y})+\frac{1}{2}(\varepsilon _{i}\sigma
_{i}^{z}+\varepsilon _{j}\sigma _{j}^{z})+J_{ij}\overrightarrow{\sigma }
_{i}\cdot \overrightarrow{\sigma }_{j}$. When turning off \ the parameters $
\varepsilon _{i},\varepsilon _{j}$ and $J_{ij}$, one gets the gate generated
by the antisymmetric term $\sigma _{i}^{y}\sigma _{j}^{z}-\sigma
_{i}^{z}\sigma _{j}^{y}$ by just waiting. Since this term is very small
compared to $J_{ij}$, to a good approximation we can neglect its effect when
we turn on other terms, e.g., $H_{ij}\approx J_{ij}(t)\overrightarrow{\sigma 
}_{i}\cdot $ $\overrightarrow{\sigma }_{j}$ when turning on $J_{ij}$. We can
then show that $SU(4)$ can be generated by commutation. E.g., $\sigma
_{i}^{y}=[[\sigma _{i}^{y}\sigma _{j}^{z}-\sigma _{i}^{z}\sigma _{j}^{y},
\overrightarrow{\sigma }_{i}\cdot \overrightarrow{\sigma }_{j}],\sigma
_{i}^{z}]/2,$ and similarly, we can generate $\sigma _{j}^{y}$. Therefore,
we have the gate set generated by $\{\sigma _{i}^{y},\sigma _{j}^{y},\sigma
_{i}^{z},\sigma _{j}^{z},\overrightarrow{\sigma }_{i}\cdot \overrightarrow{
\sigma }_{j}\}$ which is known to be universal. It is interesting to note
that the approximation assuming a small antisymmetric term is not necessary 
\cite{WuLidar:tbp}. If control over $\varepsilon _{i}$ is unavailable one
may have to resort to other methods \cite{WuLidar:tbp,Bonesteel:01}.

{\it Elimination of single-qubit operations through encoding.---} Our
discussion of universality so far assumed that one is seeking to employ the
full $2^{N}$-dimensional Hilbert space of $N$ qubits. However, it was
apparent from this discussion that the symmetries of a given Hamiltonian
determine an invariant subspace and that in physically generic circumstances
this subspace has reduced dimensionality. A common solution is to introduce
an external field which breaks the symmetry. As discussed above this often
leads to significant engineering complications . However, as shown first in 
\cite{Bacon:99a} for the case of isotropic exchange, a Hamiltonian may still
be {\em computationally universal over a subspace}, for the price of using
several physical qubits to encode a logical qubit. Here we analyze this
concept for the anisotropic members of the class of Hamiltonians~$
H_{0}+V^{\prime }$. In each case we assume that no external single qubit
operations are used, i.e., $F=0$, and give an encoded universal set of
gates. As distinct from \cite
{Bacon:Sydney,Bacon:99a,Zanardi:99dLidar:00bViola:01,Kempe:00,DiVincenzo:00a}
we explicitly take $H_{0}$ into account, as this is a term that is generally
difficult to turn off. Our analysis provides simple encoding procedures
along with explicit recipes for universal computation in situations of
experimental interest.

{\it Axial Symmetry}.--- Assume $\Delta _{ij}=0$. This axial symmetry is the
case, e.g., for the electrons floating on helium proposal
\cite{Platzman:99}. The major handle there is the single-qubit energies $\varepsilon _{i}$, which
allows to tune the qubits into and out of resonance with externally applied
radiation. This tuning is used to control the parameters $f_{i}$, $J_{ij}^{z}
$ and $J_{ij}$ of Eqs.~(\ref{eq:F}),(\ref{eq:V'}). However, it is
advantageous to do away with controlling the single qubit parameters
$f_{i}$, as they are manipulated via a global and slow microwave
field. Limitations 
related to other QC proposals were discussed above. Motivated by these
difficulties a solution involving control of only the $\sigma _{i}^{x}\sigma
_{j}^{x}+\sigma _{i}^{y}\sigma _{j}^{y}$ term was proposed in \cite
{Bacon:Sydney}, encoding a qutrit into three physical qubits. Here we give a
more economical solution: we show how to compute universally on a logical
qubit encoded into only two physical qubits. Our solution makes use of the
naturally available $H_{0}$ term, and assumes that the $J_{ij}^{z}$ and $
J_{ij}$ parameters can be tuned separately. In fact not all of these
parameters need to be independently controllable, as discussed below. Since in the axial symmetry case $V^{\prime }$
preserves occupation number, the encoding is simply $|0_{L}\rangle
_{m}=|0\rangle _{2m-1}\left| 1\right\rangle _{2m}$ and $|1_{L}\rangle
_{m}=|1\rangle _{2m-1}\left| 0\right\rangle _{2m}$ for the $m^{{\rm th}}$
logical qubit$.$ To implement single-encoded-qubit operations assume we can
selectively turn on nearest-neighbor interactions $J_{2m-1,2m}$ and $
J_{2m-1,2m}^{z}$ in pairs encoding a qubit (i.e., $
J_{2m,2m+1}=J_{2m,2m+1}^{z}=0$). Using the definitions (\ref{eq:slt}),(\ref
{eq:slr}) with $\mu \equiv m$ when $i=2m-1$ and $j=2m$, we can rewrite the
Hamiltonian (\ref{eq:H(t)}) as:

\begin{equation}
H_{{\rm AS}}=\sum_{m=1}^{N/2}\left( \frac{\epsilon _{m}}{2}
T_{m}^{z}+J_{m}T_{m}^{x}\right) +h_{1}+h_{0},
\end{equation}
where $\epsilon _{m}\equiv \varepsilon _{2m-1}-\varepsilon _{2m},$ $
J_{m}\equiv J_{2m-1,2m}$, $\omega _{m}=\varepsilon _{2m-1}+\varepsilon _{2m}$, $h_{1}\equiv \sum_{m=1}^{N/2}\frac{1}{2}\omega _{m}R_{m}^{z}$, and $
h_{0}\equiv \sum_{m=1}^{N/2}J_{2m-1,2m}^{z}\left(
(R_{m}^{z})^{2}-(T_{m}^{z})^{2}\right) $. The term $h_{0}$ is an energy
shift which commutes with all other operators, and will thus be neglected.
It is then clear that $H_{{\rm AS}}$ is a sum over independent modes $m$, so
that the Hilbert space decomposes into a tensor-product structure. The
operators $T_{m}^{z}$,$T_{m}^{x}$ generate an encoded $SU_{m}^{t}(2)$ group,
while the term $h_{1}\in su_{m}^{r}(2)$ acts as a constant (since $
[su_{m}^{t}(2),su_{m}^{r}(2)]=0$). As a whole $H_{{\rm AS}}$ acts as $
\bigotimes_{m=1}^{N/2}SU_{m}^{t}(2)$, meaning that experimental control over
the coefficients $\epsilon _{m}$ and $J_{m}$ enables the implementation of
independent and arbitrary encoded-single qubit operations. Next we need to
show how to implement an encoded controlled operation. This can be done very
simply using nearest-neighbor interactions only. All that is required is to
turn on the coupling $J_{2m,2m+1}^{z}$, since as is easily checked $\sigma
_{2m}^{z}\sigma _{2m+1}^{z}=-T_{m}^{z}T_{m+1}^{z}$. This yields a
controlled-phase gate \cite{Nielsen:book}. It may appear from this
discussion that all $\epsilon _{m}$, $J_{m}$ and $J_{2m,2m+1}^{z}$ should be
controllable. However, in analogy to NMR, we can further show that {\em 
independent control over the coefficients \ }$J_{m}${\em \ } {\em suffices
to generate arbitrary single-encoded qubit operations and an encoded
controlled operation}. Suppose that $\epsilon _{m}$ and $J_{2m,2m+1}^{z}$
are not directly controllable, as is the case for the analogous parameters
in front of the terms $\sigma _{i}^{z}$ and $\sigma _{i}^{z}\sigma _{j}^{z}$
in a typical liquid-state NMR Hamiltonian. {\em Refocusing} in terms of $
T_{m}^{x}$ then plays the same role as refocusing using $\sigma ^{x}$ in NMR 
\cite{Nielsen:book}, allowing control over $\epsilon _{m}$ and $
J_{2m,2m+1}^{z}$. This ``encoded refocusing'' method will be treated in
detail in a separate publication \cite{WuLidar:tbp}.

{\it Decoherence Avoidance.---} The connection between encoding and immunity
to decoherence is known from the theory of decoherence-free subspaces (DFSs) 
\cite{Zanardi:97cDuan:98Lidar:PRL98}. The present encoding is
decoherence-free under the following conditions: Assume that the system-bath
interaction is $H_{I}=\sum_{i=1}^{N}\sigma _{i}^{z}\otimes B_{i}^{z}$ where $
B_{i}^{z}$ are bath operators. If pairs of qubits are sufficiently close
compared to the bath wavelength, so that $B_{2m-1}^{z}=B_{2m}^{z}\equiv $ $
\tilde{B}_{m}^{z}$ (``block-collective phase damping'' \cite
{Zanardi:97cDuan:98Lidar:PRL98}) then $H_{I}\rightarrow H_{I}^{{\rm CPD}
}=2\sum_{m=1}^{N/2}R_{m}^{z}\otimes \tilde{B}_{m}^{z}$. But $R_{m}^{z}\left(
\alpha |0_{L}\rangle _{m}+\beta |1_{L}\rangle _{m}\right) =0$ so that the
interaction $H_{I}^{{\rm CPD}}$ does not cause decoherence. Furthermore, $
H_{I}^{{\rm CPD}}$ commutes with $H_{{\rm AS}}$ and with $
T_{m}^{z}T_{m+1}^{z}$, so it follows from a general theorem \cite
{Bacon:99a,Zanardi:99aKnill:99a} that with the methods provided above
universal encoded logic can be implemented without ever leaving the DFS.

{\it Axially Asymmetric Interaction}.--- Assume that one can control the
axial asymmetry parameter $\Delta _{ij}=J_{ij}^{x}-J_{ij}^{y}$ in Eq.~(\ref
{eq:V'}). Further assume only nearest-neigbor interactions in pairs are on,
and let $\Delta _{m}\equiv \Delta _{2m-1,2m}$. The Hamiltonian $
H_{0}+V^{\prime }$ now becomes:

\[
H_{{\rm AA}}=\sum_{m=1}^{N/2}\left( \frac{\epsilon _{m}}{2}
T_{m}^{z}+J_{m}T_{m}^{x}\right) +\left( \frac{\omega _{m}}{2}
R_{m}^{z}+\Delta _{m}R_{m}^{x}\right) ,
\]
where we have again omitted the $h_{0}$ term. The appropriate encoding for
the $R_{m}^{z,x}$ terms is: $|0_{L}\rangle _{m}=|0\rangle _{2m-1}\left|
0\right\rangle _{2m}$, $|1_{L}\rangle _{m}=|1\rangle _{2m-1}\left|
1\right\rangle _{2m}$ for the $m^{{\rm th}}$ logical qubit, since the
axially asymmetric component of the Hamiltonian preserves parity but not
occupation number. To implement a controlled operation on the $m^{{\rm th}
}\otimes m+1^{{\rm th}}$ encoded-qubits' Hilbert space it suffices again to
turn on the nearest neighbor coupling $\Delta _{2m,2m+1}$, since $\sigma
_{2m}^{z}\sigma _{2m+1}^{z}=R_{m}^{z}R_{m+1}^{z}.$ In analogy to the
analysis above, the subspace acted on by $su_{m}^{t}(2)$ operators is
furthermore decoherence-free if the system-bath interaction $
H_{I}=\sum_{i=1}^{N}\sigma _{i}^{z}\otimes B_{i}^{z}$ has the symmetry $
B_{2m-1}^{z}=-B_{2m}^{z}$. The two subspaces acted upon by the axially
symmetric and antisymmetric terms are independent. They can be regarded as
two independent quantum computers.

{\it State Preparation and Measurement.}--- The state $\left(
|01\rangle-|10\rangle\right) /\sqrt{2}=(|0_{L}\rangle-|1_{L}\rangle)/\sqrt {
2 }$ is the ground state of the axially symmetric Hamiltonian $\sigma
^{x}\sigma^{x}+\sigma^{y}\sigma^{y}$, while $\left( |00\rangle-|11\rangle
\right) / \sqrt{2}=(|0_{L}\rangle-|1_{L}\rangle)/\sqrt{2}$ is the ground
state of the axially antisymmetric Hamiltonian $\sigma^{x}\sigma^{x}
-\sigma^{y}\sigma^{y}$. Thus by lowering the temperature to below $J$ and $
\Delta$ (the respective strengths of the interactions), the system will
relax into the corresponding subspaces and computation can begin.
Measurement can be done in the axially symmetric case by first applying an
encoded Hadamard gate [which maps $|0_{L}\rangle\rightarrow(|0_{L}
\rangle+|1_{L}\rangle)/ \sqrt{2}$, $|1_{L}\rangle\rightarrow(|0_{L}
\rangle-|1_{L}\rangle)/\sqrt{2 }$], and then using, e.g., Kane's a.c.
capacitance scheme \cite{Kane:98}, which distinguishes a singlet from a
triplet state. In the axially antisymmetric case Kane's scheme will
distinguish the states $\left( |00\rangle \pm|11\rangle\right) /\sqrt{2}$,
so the same procedure applies.

{\it Conclusions}.--- We studied here the quantum computational power of a
generic class of anisotropic solid-state Hamiltonians. We presented simple
encodings of one qubit into two physical qubits, and schemes which enable
universal computation in the case of axially symmetric and/or antisymmetric
exchange-type Hamiltonians, while avoiding difficult-to-implement
single-qubit control terms. Only nearest-neighbor interactions are needed
for this implementation of encoded universal quantum logic. These results
can be generalized to provide codes with higher rates \cite{WuLidar:tbp}.
The methods presented here have the potential to offer significant
simplifications in the construction of QCs based on quantum dots, donor-atom
nuclear or electron spins, quantum Hall systems, and electrons floating on
helium.

{\it Acknowledgements}.--- We thank A. Blais and A. Maassen v.d. Brink
for helpful discussions. DAL gratefully acknowledges financial support
from PREA, NSERC, and the Connaught Fund.

\end{multicols}


\begin{references}
\bibitem{comment1}  We use the convention that upper-case (lower-case)
denotes a Lie group (algebra).
\end{references}
\end{document}